\documentclass[12pt]{article}

\usepackage{graphicx}
\usepackage{caption}

\begin{document}

\begin{center}

{\Large\bf 40 Years of Calculus in $4 + \epsilon$ Dimensions} \\

\vspace*{8mm}

Wolfgang Bietenholz$^{\rm a}$ and Lilian Prado$^{\rm b}$ \\
\vspace*{3mm}

{\footnotesize $^{\rm a}$ Instituto de Ciencias Nucleares \\
Universidad Nacional Aut\'{o}noma de M\'{e}xico \\
A.P.\ 70-543, C.P.\ 04510 Distrito Federal, Mexico

\vspace*{2mm}
$^{\rm b}$ Facultad de Ciencias F\'isico-Matem\'aticas \\
Benem\'erita Universidad Aut\'onoma de Puebla \vspace*{-0.5mm}\\
Puebla, Pue., Mexico }

\end{center}

\vspace*{4mm}

{\em Quantum Field Theory} is the successful formalism for particle 
physics. Analytical calculations in this framework usually
apply a {\em perturbation expansion,} {\it i.e.}\ one starts from
free particles and computes corrections, order by order
in the interaction parameter. This involves numerous divergent
terms, which need to be {\em regularized} in a subtle way, such that 
the physical limit can be taken in a controlled manner at the very end.

40 years ago a new method was established for this purpose,
known as the {\bf Dimensional Regularization}. It was the
essential tool for the revolution in theoretical particle physics 
in the early 1970s, which led to the Standard Model
that we all refer to today. However, not everybody
is aware that the origin of this regularization scheme
was in Latin America, more precisely in {\em La Plata, Argentina.}\\

In 1970 Quantum Field Theory was already established as the
appropriate formalism for elementary particles. This was based on the
success of Quantum Electrodynamics (QED), where electrically
charged particles interact through the Abelian gauge group $U(1)$.
The idea to extend gauge theory to non-Abelian Lie groups, like $SU(2)$, 
was also known, and denoted as {\em Yang-Mills theories.}
In particular, this was required for the proposal to unify 
electromagnetic and weak interactions \cite{SM}. However, such theories 
seemed to be ``non-renormalizable''; it was not known how to compute
finite and predictive results. The regularization
schemes that were used at that time to handle the divergences 
(like the Pauli-Villars method) failed for Yang-Mills theories.

However, in 1971 Gerard 't Hooft, a brilliant Ph.D.\ student in 
Utrecht (Netherlands), formulated a program of how to make sense 
out of such theories, if a suitable regularization scheme would 
be available \cite{tHooft71}. In particular, gauge invariance should 
be preserved on the regularized level. Meanwhile physicists thought 
about new schemes in many countries, including {\em Argentina}. \\

A remarkable development had taken place at the Universidad
de Buenos Aires (UBA) in the late 1950s: a young professor named
{\bf Juan Jos\'{e} Giambiagi} became the head of the Physics Department from 
1957 to 1966; under his direction it expanded, gained
worldwide reputation and attract highly motivated students. Among 
them was M.A.\ Virasoro, who is now famous for the {\it Virasoro
Algebra.} Together with his long-term collaborator {\bf Carlos
Guido Bollini}, Giambiagi was particularly interested in applications
of {\em distributions} (or {\em generalized functions}) to 
particle physics. They studied thoroughly the books 
by I.M.\ Gel'fand and G.E.\ Shilov, in particular
the first volume \cite{GelShi}. Still in his older age Giambiagi called 
it ``my bible'', and he was convinced that it contained a great potential 
for the physics of the future. In 1964 Bollini and Giambiagi --- together
with A.\ Gonz\'{a}lez Dom\'{\i}nguez, a mathematician who had
supervised Giambiagi's Ph.D.\ thesis  --- suggested a new scheme that 
they called Analytic Regularization, which works in QED \cite{anareg}.
Ref.\ \cite{Speer72} discusses mathematical aspects of the
renormalization in this scheme.

In 1966 a {\it coup d'\'{e}tat} brought General Ongan\'{\i}a to power.
He stopped university autonomy, which had been recognized in Argentina 
since 1919. In protest, students and professors occupied five faculties 
of UBA, which were violently cleared by the Federal Police, arresting 
hundreds of persons and destroying laboratories and libraries, in the 
``Noche de los Bastones Largos'' (July 29).
The subsequent repression led to the resignation of
about 1500 lecturers \cite{bio}. Also the excellent Physics Department
fell into ruin, despite a public letter of support, which was signed
by 192 renowned physicists, including six Nobel laureates
at that time, and eight future ones.
\begin{figure}[h!]
\centering
\includegraphics[angle=0,width=0.5\linewidth]{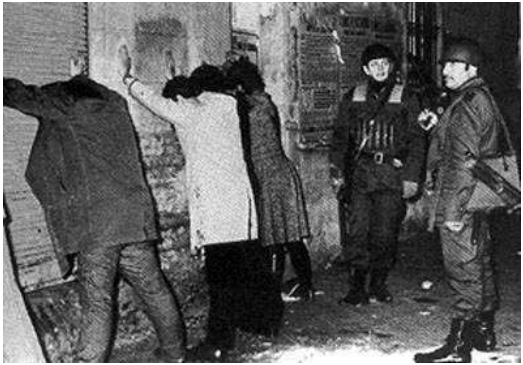} 
\hspace*{3mm}
\includegraphics[angle=0,width=0.45\linewidth]{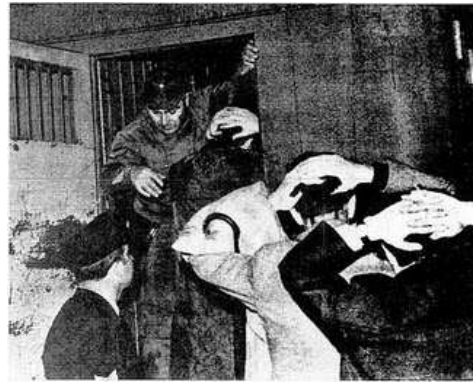}
\caption*{Pictures of the {\it Noche de los Bastones Largos}
(``Night of the Long Batons''), Universidad de Buenos Aires, 
July 29, 1966.}
\end{figure}

Bollini and Giambiagi moved south-east to La Plata, where
they worked from 1968 to 1976. Despite difficult working 
conditions, it turned out to be their most productive period.
Gifted students from UBA joined, among them F.\ Schaposnik, who 
is now a renowned leader in Argentinian high energy physics. In 
1971 Bollini and Giambiagi came up with a new idea of how to regularize
the divergences in perturbative Quantum Field Theory, which truly
surprised many people: they calculated in $\, 4 + \epsilon$ {\em space-time
dimensions.} They arrived at finite results by analytic continuation
in $\epsilon$ all over the complex plane, up to poles at 
$\epsilon = -2, \, 0, \, 2, \, 4 \dots $ . The limit $\epsilon \to 0$
should be taken at the very end. They first applied this method
to scalar (spin-less) particles. In November 1971 they submitted
an article to the well-reputed journal {\it Physics Letters B,} which
belongs to the Dutch publishing house {\it Elsevier.} 
However, the editors and referees found this approach
too surprising and blocked its publication. Giambiagi commented 
later ironically that it was rejected ``because we didn't
know that the space-time dimension is 4''.

Hence Bollini and Giambiagi wrote another article about their
new approach, now more extensive and with applications to QED,
where they demonstrated gauge invariance.
They sent it to the Italian journal {\em Il Nuovo Cimento B,} 
where it arrived on February 18, 1972. There it was accepted, 
but it took until November to get published \cite{NC72}.
\begin{figure}[h!]
\centering
\includegraphics[angle=0,width=0.43\linewidth]{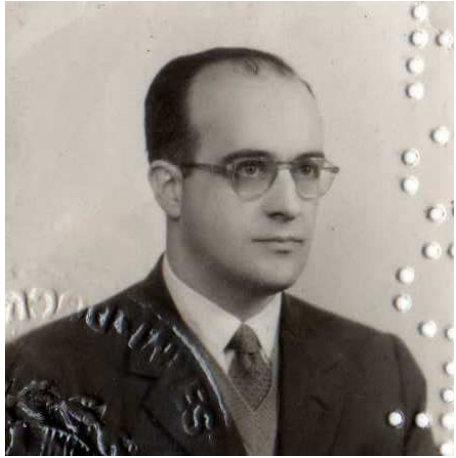} \hspace*{7mm}
\includegraphics[angle=0,width=0.33\linewidth]{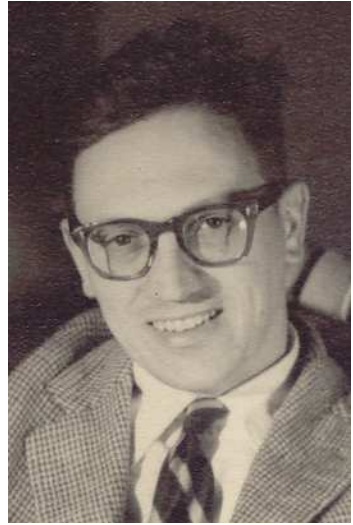}
\caption*{Carlos Guido Bollini (1926 - 2009, on the left) and Juan
Jos\'{e} Giambiagi (1924 - 1996, on the right), the two Argentinian
physicists who invented Dimensional Regularization in 1971.}
\end{figure}

Meanwhile, on February 21, 't Hooft and his Ph.D.\ advisor
Martinus Veltman submitted an article to {\em Nuclear Physics B}
(another journal that belongs to {\em Elsevier}), 
which suggested the same regularization procedure. 
Their work was more extensive, and it
contained applications to Yang-Mills theories, which can be
successfully renormalized in this way. That article was published
swiftly, on July 1 \cite{NPB72}. About a month later also the original 
Argentinian work finally appeared in {\it Physics Letters B} \cite{PLB72}.

Thus the Dutch paper was first to be published, due to the
handling by the editors, although it was submitted later.
It is generally viewed as an independent work, a little later 
than the one from Argentina, but much more comprehensive. 
However, it is a matter of fact that 't Hooft and Veltman already
knew the second article from La Plata, since they quoted it as a 
preprint. A further delicate detail is that Veltman was a member of the
Advisory Editorial Board of {\em Physics Letters B} during that time.\\

Now Dimensional Regularization led to a 
revolution in particle physics: 
the {\em electroweak unification} was shown to work, and it 
was also confirmed experimentally: in 1973 CERN observed the
neutral weak current that had been predicted, and later it also 
identified all three massive weak gauge bosons. Also in 1973
{\em Quantum Chromodynamics} (QCD) was elaborated 
\cite{FGML}. It is again based on a 
Yang-Mills theory, this time with the gauge group $SU(3)$. The strong 
interaction is transmitted by 8 gluons, corresponding to
the 8 generators of $SU(3)$, and they interact among each other,
as the non-commutativity of these generators reveals. In the same
year, the QCD property of {\em asymptotic freedom} was understood
\cite{GWP}, 
which led to a perfectly consistent picture of the hadronic world.
Thus the Standard Model of particle physics was established.
It holds to high precision to the present (if neutrino masses are
incorporated), and it is likely that CERN has just discovered 
the Higgs particle as its last ingredient.\\

On the other hand, in 1976 Argentina suffered from another military 
{\it coup} and the beginning of the Dirty War. This also affected
the Universidad de La Plata: Giambiagi was interrogated by the
Federal Police about links to an alleged Jewish or communist
conspiracy \cite{bio}. He and Bollini both escaped to Rio de Janeiro, 
where they soon worked at the {\em Centro Brasileiro de Pesquisas
F\'{\i}sicas} (CBPF). Giambiagi was the head of its Particle
Physics Department from 1978 to 1985. One year later
he became director of the Centro Latinamericano de F\'{\i}sica 
(CLAF), which he had founded back in 1960 with two prominent colleagues
from Brazil and Mexico: Jos\'{e} Leite Lopes and Marcos Moshinsky.
Giambiagi encouraged young scientists at CLAF to address also
subjects of practical relevance, such as energy resources and
climatic issues \cite{bio}.
He stayed in Rio de Janeiro until he passed away in 1996. Bollini
returned to Argentina --- after the military regime had ended ---
where he lived until 2009.
\begin{figure}[h!]
\centering
\includegraphics[angle=0,width=0.43\linewidth]{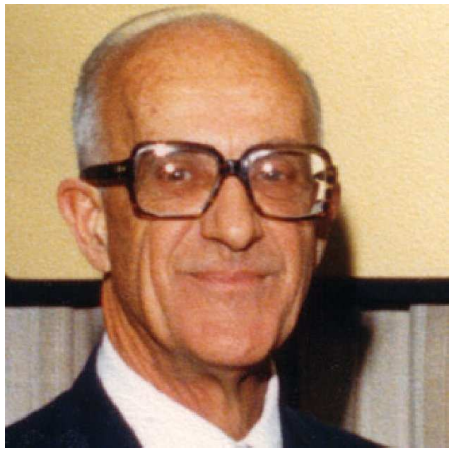} 
\hspace*{1cm}
\includegraphics[angle=0,width=0.33\linewidth]{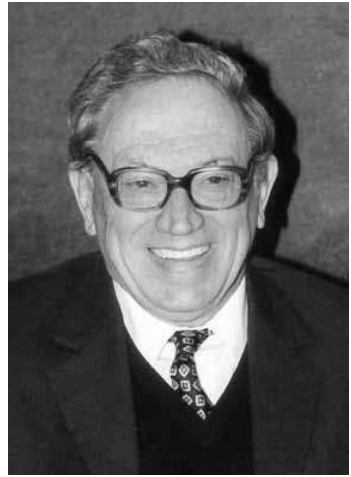}
\caption*{Carlos Guido Bollini (on the left) and Juan
Jos\'{e} Giambiagi (on the right), after their escape to Brazil.}
\end{figure}
\\

In 1999 't Hooft and Veltman received the Physics Nobel Prize. 
In particular 't Hooft certainly deserved it for numerous important
works. However, the official press release only referred 
to four papers: the most cited among them is the one where 
Dimensional Regularization was re-invented and applied to Yang-Mills 
theories \cite{NPB72}. The Swedish Royal Academy mentioned the names 
of 15 other researchers who had done related work, but it ignored 
the Argentinian contribution \cite{ceci}. Only
the Nobel lecture by Veltman contains a side-remark about
``the independent work of Bollini and Giambiagi'', though
it is not included in his list of references \cite{velt}.
\begin{figure}[h!]
\centering
\includegraphics[angle=0,width=0.5\linewidth]{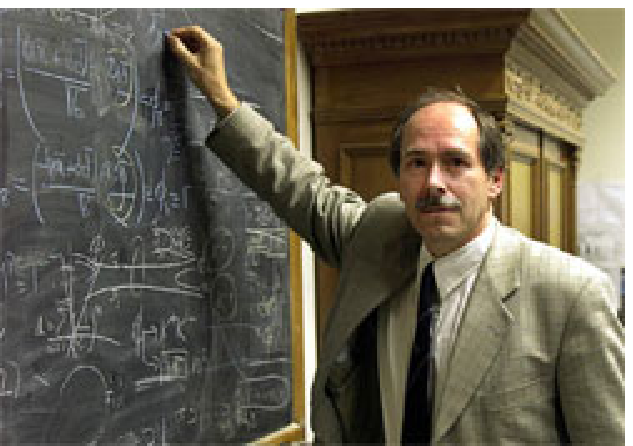}
\hspace*{6mm}
\includegraphics[angle=0,width=0.4\linewidth]{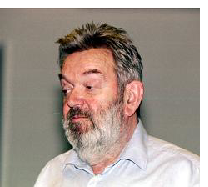}
\caption*{Gerard 't Hooft (on the left) and Martinus Veltman
(on the right), Nobel Prize winners of 1999, ``for elucidating 
the quantum structure of electroweak interactions in physics''.
}
\end{figure}

To this day, the paper by 't Hooft and Veltman 
on Dimensional Regularization has 2929 citations (according
to http://inspire.net/ ), whereas the two articles from La Plata
have 298 and 579 citations, respectively. This is also considerable,
but only little compared to the immense impact of Dimensional 
Regularization, which is now the standard method in 
perturbative calculations, explained in the Quantum Field Theory
text books, and which was essential for the breakthrough of the 
Standard Model. 

It still happens often that papers and seminar speakers refer to 
this scheme, quoting only 't Hooft and Veltman --- also here in Mexico.
We hope for this article to counteract a little. At last we congratulate 
the  Dimensional Regularization to its 40 years of existence as
a highly successful procedure --- and 41 years since it was invented
in La Plata.\\

\noindent
{\it It is a pleasure to thank Ernesto Bollini, Marcia Giambiagi, 
Jos\'{e} Helay\"{e}l-Neto, Mario Rocca and Fidel Schaposnik for 
helpful communications.}

\newpage

\end{document}